# BER Analysis of Full Duplex Relay assisted BPSK-SIM based VLC System for Indoor Applications

L Bhargava Kumar[1], Ramavath Prasad Naik[2,*], Datta Choudhari[3], Prabu Krishnan[4], Goutham Simha G D[5], and Jagadeesh V K[6]

[1]School of Electronics and Communication Engineering, KLE Technological University, Hubballi, Karnataka, India
[2]Department of Electronics and Communication Engineering, National Institute of Technology Rourkela, India
[3]School of Electronics Engineering, Vellore Institute of Technology, Vellore, India
[4]Department of Electronics and Communication Engineering, National Institute of Technology Karnataka, Surathkal, India
[5]Department of Electronics and Communication Engineering, Manipal Institute of Technology, Manipal Academy of Higher Education (MAHE), Manipal, Karnataka, India.
[6]Department of Electronics and Communication Engineering, National Institute of Technology Patna, Bihar, India

Email: * naikrp@nitrkl.ac.in

**ABSTRACT-** This paper contemplates a relay-assisted visible light communication (VLC) system, where the light source (Table lamp) acts as a relay node and cooperates with the main light source. Following the IEEE 802.15.7r1 VLC reference channel model, we assume that there are two different light sources present in an office room. The first one is the source terminal present on the ceiling and another one is the desk lamp that serves as the relay station which works in full-duplex method. Because of the loop interference channel, we model VLC relay terminal using ray tracing simulations. We have analyzed bit error rate (BER) performance of the relay-assisted VLC system using binary phase shift keying-subcarrier intensity modulation (BPSK-SIM) technique. The proposed method outperforms existing phase shift keying (PSK) and square M-quadrature amplitude modulation (M-QAM) techniques. The proposed VLC system using BPSK-SIM technique achieves a BER performance of $10^{-12}$ for an SNR of 20 dB. The results of proposed full duplex and half duplex relayed VLC system are evaluated using equal power allocation (EPA) and optimum power allocations (OPA) techniques over three different modulation schemes which are 2-PSK, square M-QAM, BPSK-SIM.

## 1. Introduction

Visible light communication (VLC) is a data communications alternative which uses visible light over a wavelength of 375-780 nm. VLC has come into view as a modified short-range wireless transmission technique also known as Li-Fi using light emitting diodes (LEDs) which is used for illumination purposes as well [1]. Since the visible spectrum range is from 375-780 nm and RF spectrum range varies from 20 kHz – 300 GHz, it is quite impossible to interface VLC with RF. As the RF spectrum is almost impregnated, VLC is the best alternative technique to RF. In the current illumination infrastructure for data communication purposes, VLC put forward an energy-efficient and low-cost solution. For transmitting data, VLC uses LEDs and for receiving the light, photo detectors (PDs) are used. Nevertheless, bipolar or complex signals cannot be transmitted through LEDs, because of its non-coherent properties. As a consequence of that, intensity modulation and direct detection (IM/DD) method is widely preferred in VLC systems.

Specifically, VLC can be used in some indoor environments such as office rooms, living room, workplace, and sports complex with the involvement of secondary light. The secondary lights present in indoor environment, provide motivation to use them as a relay terminal between source and destination to increase the link reliability which in turn increases the coverage area [2]-[4]. Either full-duplex (FD) or half-duplex (HD) relaying can be utilized. Equivalently, we can model the main source i.e., ceiling light as a relay terminal when there is more than one ceiling light sources present at few meters away from each other for enhanced coverage. In radio communication environment most often HD relaying is a preferred choice over FD, considering the fact that in FD relaying, there exists an adequate amount of difference in transmit and received power levels. Due to the coupling of strong transmitted signal with receiver and overpowering of the weaker received signal in a FD radio relay station, the loop interference is at a greater level [5]. In VLC, it is easy to implement FD relaying as their loop interference is at lower levels because of the concentrated illumination of LEDs. Initially the work started in VLC field on simple modulation techniques such as pulse position modulation (PPM) and on-off-keying (OOK) [6]. Though, inter-symbol interference (ISI) is one of the major challenges in VLC because of its multi-path nature, targeting data rates on the order of multiple Gbps over a multi-path VLC channel causes typical delay spreads in the order of several nanoseconds and additionally this introduces ISI. This ISI effect can be reduced using time-domain equalization in multi-carrier communication techniques or in single-carrier systems, by inducing the concept of orthogonal frequency-division multiplexing (OFDM).

In [8, 9], authors have analyzed the performance of relay-assisted OFDM based VLC system and compared with the direct transmission schemes. That is, by incorporating HD relaying technique, the performance of VLC system can be improved significantly with respect to a direct transmission system. As the modulation size increases the gain will be reduced to a greater extent. The spectral efficiency of FD relaying system is good compared with HD because of its additional time slot usage for a relaying phase.

The quality of data transmission in a VLC system can be diminished by the presence of obstacles between the source and destination terminals. In most cases, multiple illuminators are present in an indoor room environment. We can use one of the illuminators as a relay terminal to help transmit data between the source and destination terminals. In this regard, a VLC system can be combined with a relaying technique to improve communication stability and indoor coverage. A cooperative VLC system is studied with FD and HD relaying modes and the results are compared in [15]. Full-Duplex (FD) and Half-Duplex (HD) relay applications in VLC systems are investigated extensively [19].

In this work, FD relay assisted VLC system using OFDM has been investigated. Firstly, an indoor VLC channel model environment is developed and analyzed using its channel impulse response (CIR) from relay transmitter to relay receiver. Our contributions in this paper can be précised as follows:

(1) We analyze the error rate performance of VLC system using relaying techniques over IEEE 802.15.7r1 reference channel model as given in [10]. Further low pass filter characteristic of LEDs is considered for evaluating different modulation techniques i.e., 2-PSK, square M-QAM and BPSK subcarrier intensity modulation (BPSK-SIM).

(2) We introduce optimum power allocation (OPA) between sources and relay terminals to enhance the bit error rate (BER) performance for the same and the results reveal that FD relaying completely outperforms the HD relaying scheme.

The remaining paper is organized as follows: an FD indoor VLC channel model with loop interference is described in Section 2. The overall system model is described in Section 3. The BER analysis of the proposed system is explained in Section 4. Section 5, describes the numerical results and discussions, followed by conclusions in Section 6.

## 2. Full Duplex (FD) indoor VLC channel model

The IEEE has built up the standardization group 802.15.7r1 "Short Range Optical Wireless Communications" (SROWC) which is right now under development of becoming a standard for VLC. In progression for standardization work, IEEE 802.15.7r1 was proposed by IEEE for assessment of VLC framework proposition. These models were created for ordinary indoor environments which involve manufacturing cells, office, and home [10-12]. In this proposed work, we have considered an indoor office environment with two light sources (ceiling light and desk light) as shown in Fig 1. The ceiling light acts as a source (S) and desk light acts as a relay terminal (R). The USB receiver is located next to the computer on the desk acts as a destination (D). The proposed system is especially useful for assessing the performance of relay-assisted (cooperative) VLC systems, in which the ceiling light serves as the source and the desk light serves as the relay. The destination receiver is next to the laptop on the desk. This could, for example, be a USB device connected to the laptop. The relay receiver is positioned on top of the desk light, with a 45° rotation towards the ceiling source. The room's dimensions are 5 m X 5 m X 3 m.

In this work, half-duplex (HD) and full-duplex (FD) relaying is considered. FD relay represents source to the destination and HD relay represents a communication between either source to relay or relay to destination. Figure 1(a) shows the full duplex relaying scenario between the source (S) and destination (D) via the relay (R). Figures 1(b) and 1(c) show the half duplex relaying scenarios from source (S) to relay (R) and relay (R) to destination (D), respectively.

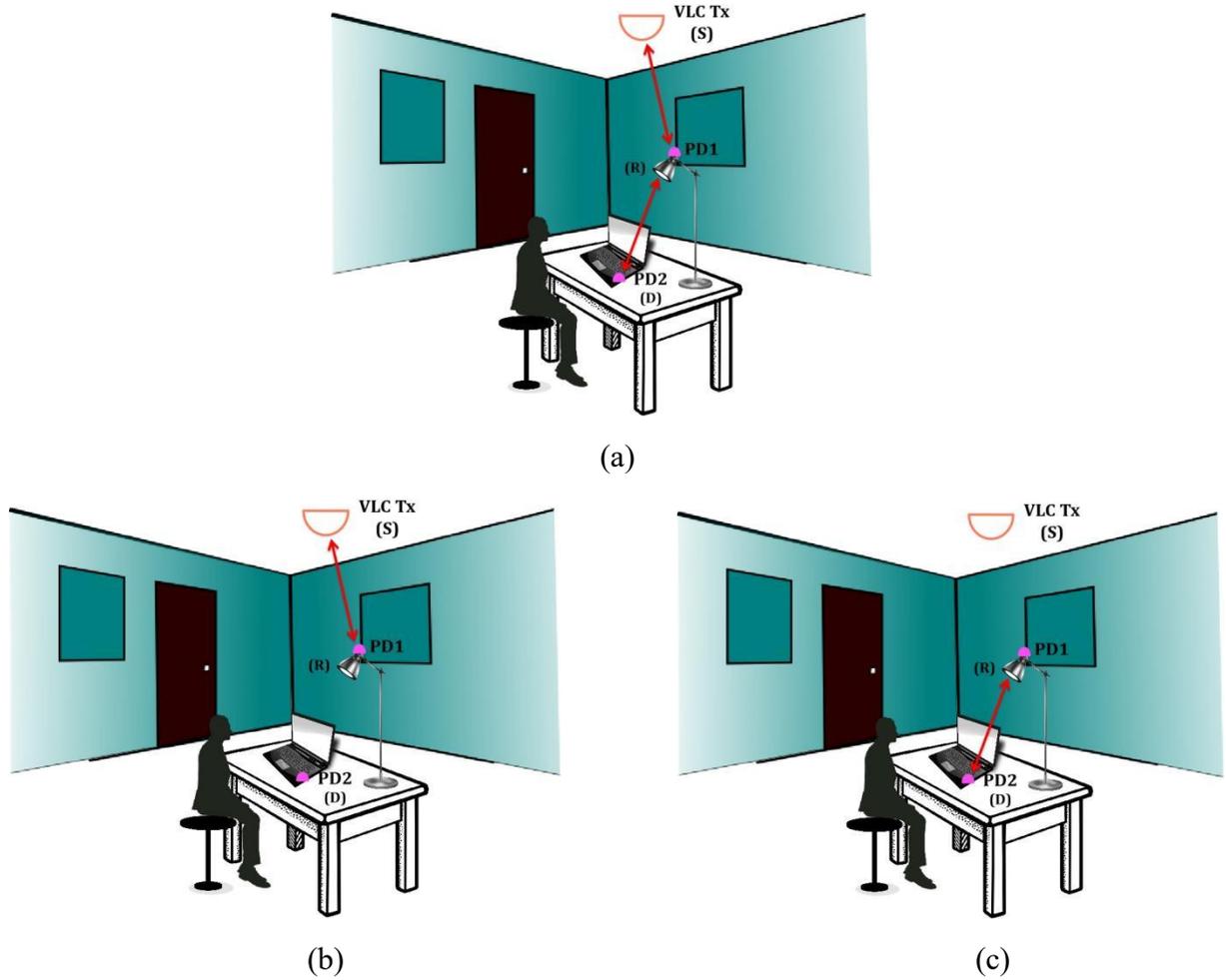

(a)

(b)                                    (c)

Fig. 1. VLC Indoor environment (a) Full duplex relaying (b) Half duplex relaying from S to R (c) Half duplex relaying from R to D

The optical CIR from source to destination, source to relay and relay to destination are respectively denoted as $C_{sd}$ (t), $C_{sr}$ (t) and $C_{sd}$ (t). CIRs for source-to-destination, source-to-relay and relay-to-destination are provided in [10]. Furthermore, the effects of LEDs are considered in the channel modeling due to the multi-path propagation characteristics. The frequency response of LED is presumed to be [13]

$$C_{LED}(f) = \frac{1}{1 + j(\frac{f}{f_{cut-off}})} \qquad (1)$$

Where $f_{cut-off}$ is 3-dB cut-off frequency of LED. Assuming $f_{cut-off}$ = 20 MHz. It can be observed that attenuation has been increased over the higher frequencies because of low pass characteristics of the LEDs. In our case, we consider a channel modeling methodology presented for scenario 2 in [11], in which objects such as human body, furniture; and wavelength-subordinate reflection attributes of surface materials (floor, roofs, dividers, furniture, attire) are present. We accomplish CIR (denoted as $C_{rr}$(t) associates with loop interference channel (R - R) i.e., from desk light LED to its own PD-receiver using non-sequential ray tracing approach. Simulation parameters considered to evaluate the proposed system are a LED source of $40^o$ half view angle, field of view and surface area of PD are $85^o$ and 1 $cm^2$, respectively, and indoor channel is realized by considering walls and ceiling are plaster, floor and desk are made up of pinewood, desk light, Shoes and laptop are coated with black gloss paint, hands and head are absorbing and clothes are cotton which are given in Table 1 [15].

TABLE. 1 Indoor Office Room Specification [15]

| Luminary Specifications | Brand : LR24-38SKA35 Cree Inc. Half viewing angle: $40^o$ |
|---|---|
| Field of view and Surface area of Photodetector | $85^o$ and 1 $cm^2$ |
| Stipulations of Surfaces and Objects | Walls and Ceiling: Plaster <br><br> Floor and Desk: Pinewood <br><br> Desk light, Shoes and Laptop: Black gloss paint <br><br> Hands and head: Absorbing <br><br> Clothes: Cotton |
| Transmit power | -30 dBm |
| Area of room | 5 m X 5 m X 3 m |

Due to multi-path propagation and low-pass filter characteristics of LED, the effective CIR $C_{sd-eff}$(t), $C_{sr-eff}$(t) and $C_{rd-eff}$(t) and $C_{rr-eff}$(t) respectively denote the effective CIR for source to destination (S - D), source to relay (S - R) and relay to destination (R - D) and loop interference is expressed as $C_{sd-eff}(t) = C_{sd}(t) \otimes C_{LED}(t)$, $C_{sr-eff}(t) = C_{sr}(t) \otimes C_{LED}(t)$, $C_{rd-eff}(t) = C_{rd}(t) \otimes C_{LED}(t)$ and $C_{rr-eff}(t) =$ $C_{rr}(t) \otimes C_{LED}(t)$ where, $C_{LED}(t) \overset{\leftrightarrow}{FT} C_{LED}(f)$. The corresponding effective CIRs are shown in Fig. 2, and it is examined that low-pass filter characteristics of LED are the main source of frequency selectivity.

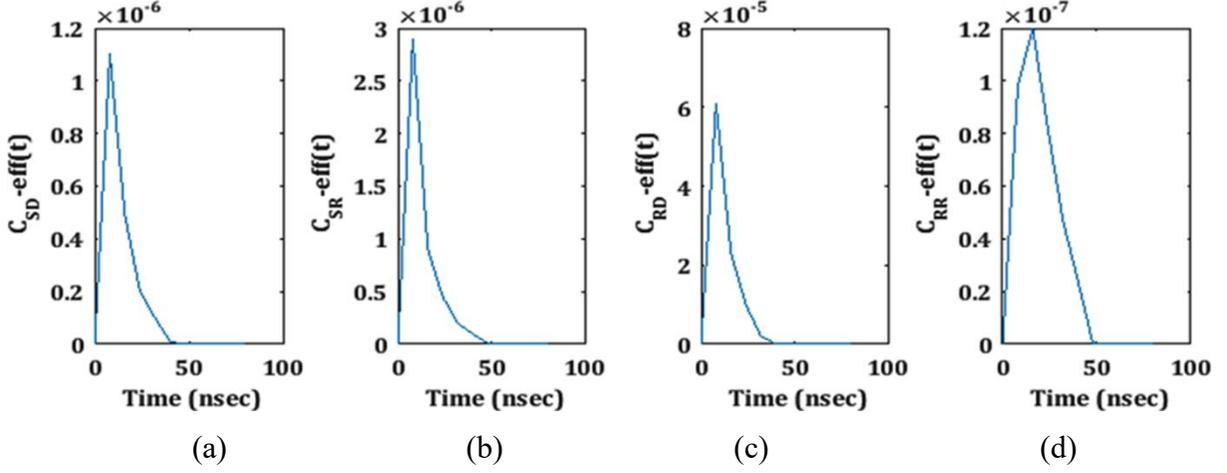

Fig. 2. Effective CIR (a) $S \to D$ (b) $S \to R$ (c) $R \to D$ (d) $R \to R$ links

The mathematical model of amplify and forward (AF) based FD relay terminal is described in [14]. In FD relay assisted transmission, relay terminal receives the information from source (S), amplify the received signal with a scaling factor $G_A$ and forward it to the destination. The average power transmitted from source to the relay is determined as $P_{Source, Tx} = PK_p$ and $P_{relay, Tx} = P(1 - K_p)$. The average power received at the relay is expressed as

$$P_{relay, Rx} = P_{Source, Tx} r^2 \int |C_{sr-eff}(t)|^2 + P_{relay, Tx} r^2 \int |C_{rr-eff}(t)|^2 + \sigma_v^2 \qquad (2)$$

Where, $P$ represent the electrical power budget between the sources and relay terminals. $K_p$ is the optimization parameter, $r$ is the responsivity of PD and $\sigma_v^2$ is variance for Additive White Gaussian Noise (AWGN) having $N_o$ as the spectral density. The terms given in Eq. (2) denote the received signal power from source and relay respectively.

CIR for FD relay is separately represented for signal and noise inputs, and is given as [15]

$$h_{FD,signal}(t) = G_A(r\delta(t - T_p)) \otimes C_{LED}(t) + r\delta(t - T_p) \otimes h_{FD,signal}(t) \otimes C_{rr-eff}(t) \quad (3)$$

$$h_{FD,noise}(t) = G_A(r\delta(t - T_p)) \otimes C_{LED}(t) + r\delta(t - T_p) \otimes h_{FD,noise}(t) \otimes C_{rr-eff}(t) \quad (4)$$

Similarly, the impulse response for HD relaying is calculated using square root raised cosine pulse shape filter. The band-limited electrical channel response for S $\to$ D, S $\to$ R and R $\to$ D links are given as [16],

$$h_{sd}(t) = g_T(t) \otimes C_{sd-eff}(t) \otimes g_R(t) \quad (5)$$

$$h_{sr}(t) = g_T(t) \otimes C_{sr-eff}(t) \otimes g_R(t) \quad (6)$$

$$h_{rd}(t) = g_T(t) \otimes C_{rd-eff}(t) \otimes g_R(t) \quad (7)$$

Where, $g_T(t)$ and $g_R(t)$ are the electrical transmit and receive pulse shaping filters respectively and

$$G_A = \sqrt{\frac{2(1-K_p)P_{Source,\,Tx}}{2K_p P_{Source,\,Tx} r^2 \int |h_{sr}(t)|^2 + \sigma_v^2}}$$

Where $G_A$ is the amplification factor.

## 3. VLC System Model

A DC-biased optical (DCO)-OFDM based VLC system is designed in this section. Complex modulation techniques such as phase-shift keying (PSK) and quadrature amplitude modulation (QAM) are considered. The input bit streams are translated at the source terminal into complex valued symbols $S$ in accordance with the employed modulation methods. Then the input vector $X$ is passed through the inverse DFT (IDFT) block and to ensure that the output is real-valued and non-negative, Hermitian symmetry is enacted which gives

$$\mathbf{X} = [0 \; S_1 \; S_2 \; ... \; S_{\frac{N}{2}-1} \; 0 \; S_{\frac{N}{2}-1}^* \; ... \; S_2^* \; S_1^*] \qquad (8)$$

where N is the number of subcarriers.

As we have discussed, the output of the IDFT block can be obtained as

$$x[n] = \frac{1}{\sqrt{N}} \sum_{k=0}^{N-1} X[k] e^{j2\pi nk/N}, \quad n \in \{0, 1, ..., N-1\} \qquad (9)$$

Where, $X[k]$ is the $k^{th}$ element of $X$. Let n = 0,1, ...$N + N_{cp} - 1$ represents the cyclic prefix (CP) results in length $N_{cp}$ which reduces inter-symbol interference (ISI). Hence, the real valued waveform in continuous time domain that drives the LED can be written as

$$x[t] = \sum_{n=0}^{N+N_{cp}-1} x[n]\delta(t - nT_s) \qquad (10)$$

Where $\delta(.)$ is the Dirac delta function and $T_s$ is the symbol length.

The input signal emitted from the source reaches the destination through a relay terminal and is termed as three node dual hop communication process. The relay terminal merely receives the information from the source and transmits it to the destination without having access to the backbone network. The CIR that includes information about both hops can be expressed as follows.

$$h_{signal}(t) = C_{sd-eff}(t) + C_{sr-eff}(t) \otimes h_{FD,signal}(t) \otimes C_{rd}(t) \qquad (11)$$

$$h_{noise}(t) = h_{FD,noise}(t) \otimes C_{rd}(t) \qquad (12)$$

After matched filtering the received electrical signal at the destination can be written as

$$y_D(t) = \left\{ \sqrt{PK_p}\, r\, x[t] \otimes h_{signal}(t) + r\, v_R(t) \otimes h_{noise}(t) + v_D(t) \right\} \otimes g_R(t) \qquad (13)$$

Where $V_D(t)$ is AWGN with mean and variance of zero and $\sigma_v^2$ respectively. Now, sampled version of $h_{signal}(t)$ can be written as

$$h_{signal}[n] = h_{signal}(t)\big|_{t=nT_s} \qquad (14)$$

The equivalent frequency domain signal at $k^{th}$ subcarrier can be written as

$$Y_D[k] = \sqrt{PK_p}\, r\, X[k] H_{signal}[k] + r\, V_R[k] H_{noise}[k] + V_D[k] G_R[k] \qquad (15)$$

Where, $h_{signal}[n] \overrightarrow{DFT} H_{signal}[k]$ , $v_R[n] \overrightarrow{DFT} V_R[k]$ , $v_D[n] \overrightarrow{DFT} V_D[k]$ , $g_R[n] \overrightarrow{DFT} G_R[k]$ , and $h_{noise}[n] \overrightarrow{DFT} H_{noise}[k]$. The final decision taken based on Maximum Likelihood (ML) can be expressed as

$$\hat{X} = \arg min_{x=\varphi} \left( \left\| Y_D[k] - \sqrt{PK_p}\, r\, X\, H_{signal}[k] \right\|^2 \right) \qquad (16)$$

Where, $\varphi$ represents the constellation points of the modulation scheme used.

## 4. BER Analysis

## 4.1 BER Vs SNR Analysis

In FD relay assisted system SNR per subcarrier based on (16) can be calculated as

$$SNR_{FD}[k] = \frac{PK_p r^2 |H_{signal}[k]|^2}{\sigma_v^2 \left(1 + |rG_A H_{noise}[k]|^2\right)} \qquad (17)$$

Similarly, SNR per subcarrier for HD relaying can be written as

$$SNR_{HD}[k] = \frac{2PK_p r^2 |H_{SD}[k]|^2}{2\sigma_v^2} + \frac{2PK_p r^4 G_A^2 |H_{SR}[k] H_{RD}[k]|^2}{2\sigma_v^2 \left(1 + |rG_A H_{RD}[k]|^2\right)} \qquad (18)$$

The average BER expression amongst the subcarrier can be calculated by

$$\text{BER} = \frac{2}{N-2} \sum_{k=1}^{\frac{N}{2}-1} BER_{SC}[k] \qquad (19)$$

Where, $BER_{SC}[k]$ given in (19) denotes the BER per subcarrier in terms of SNR and it can be analyzed for different modulation schemes as shown in [17].

The system's BER performance may be increased by selecting the right value of $K_p$, which stands for the ideal amount of power distribution between the source and relay terminal. Basically, $K_p$ is used to calculate the BER minimization for cooperative systems with FD and HD relaying. The maximization of SNR in Eq (17) is directly related to the reduction of BER per subcarrier in (19) due to negative exponential dependence. With the aid of the brute-force technique, the BER performance of the proposed VLC system is enhanced. The optimum value of $K_p$ is given in Table 1 reproduced from [15]. In this work, it can be noticed that the optimum $K_p$ is 0.8118 for P = 1 dBm.

$$BER_{SC}[k] = \begin{cases} \frac{1}{2} \text{ erfc } \left( \sqrt{SNR[k]} \right), & \text{BPSK} \\ \frac{(\sqrt{M}-1)}{\sqrt{M}log_2\sqrt{M}} \text{ erfc } \left( \sqrt{\frac{3SNR[k]}{(2(M-1))}} \right), & \text{square M} - \text{QAM} \\ \frac{1}{2} \text{ erfc } (SNR[k]), & \text{BPSK} - \text{SIM} \end{cases} \qquad (20)$$

## 4.2 BER Vs SINR Analysis

The average signal to interference plus noise ratio (SINR) over the communication floor is given as [18]

$$\overline{SINR} = \frac{1}{N_T} \sum_i \sum_j SINR_{i,j}$$

Where $N_T$ is the number of test points over the communication floor, and $SINR_{i,j}$ is the signal to interference plus noise ratio for an $i^{th}$ source and the $j^{th}$ receiver, which is given as

$$SINR_{i,j} = 10 \ log \left[ \frac{(r \ P_{i,j}^{Sig})^2}{\eta \ B + (r \ P_{i,j}^{Intf})^2} \right]$$

Where $P_{i,j}^{Sig}$ and $P_{i,j}^{Intf}$ are the signal strength and interference in the optical domain. The parameters '$r$' represents the responsivity of the photo detector, '$\eta$' represents the noise power spectral density, and '$B$' represents the LED's modulation bandwidth. The average signal strength and interference over a communication floor are given in terms of its signal strength and interference as follows.

$$\overline{P^{Sig}} = \frac{1}{N_T} \sum_i \sum_j P_{i,j}^{Sig}$$

$$\overline{P^{Intf}} = \frac{1}{N_T} \sum_i \sum_j P^{Intf}_{i,j}$$

The signal strength and the interference are given as

$$P^{Sig}_{i,j} = P^{LOS}_{i,j} = H^{LOS}_{i,j} \times P_{Ti}$$

$$P^{Intf}_{i,j} = P^{ISI}_{i,j} + P^{CCI}_{i,j}$$

$$P^{ISI}_{i,j} = P^{NLOS}_{i,j} = \sum_{ref} H^{NLOS}_{i,j} \times P_{Ti}$$

$$P^{CCI}_{i,j} = \sum_{k \neq i} H_{k,j} \times P_{Tk}$$

$$H_{k,j} = H^{LOS}_{k,j} + H^{NLOS}_{k,j}$$

TABLE 2. Optimum values of $K_p$[15]

| Average Signal Power (P in dBm) | $K_p$ | Average Signal Power (P in dBm) | $K_p$ |
|---|---|---|---|
| 0 | 0.8019 | 10 | 0.9306 |
| 1 | 0.8118 | 11 | 0.9306 |
| 2 | 0.8316 | 12 | 0.9405 |
| 3 | 0.8415 | 13 | 0.9504 |
| 4 | 0.8613 | 14 | 0.9504 |
| 5 | 0.8712 | 15 | 0.9603 |
| 6 | 0.8811 | 16 | 0.9702 |
| 7 | 0.8910 | 17 | 0.9702 |
| 8 | 0.9108 | 18 | 0.9702 |
| 9 | 0.9207 | 19 | 0.9801 |

**5. Numerical results and Discussions:**

In this section, the BER analysis of the proposed FD relay assisted VLC system is discussed. The results are compared with the HD relaying scheme under both equal power allocation (EPA) and optimum power allocations (OPA) using three different modulation schemes which are 2-PSK, square M-QAM, BPSK-SIM. Simulation parameters considered to evaluate the proposed system are: $N=256$, $N_{cp}=32$, $T_s=250$ nsec, $r=0.28$ A/W, $T_p=50nsec$, 0.5 rolling factor of pulse shaping filters ($g_T(t)$ and $g_R(t)$) and noise power spectral density of $10^{-20}$ W/Hz [15] and are given in Table 3. In the EPA method, the power of total information signal is shared between source and the relay equally i.e. $K_p = 0.5$. Similarly, in OPA, optimized value of $K_p$ for our analysis is 0.8118.

TABLE 3 Simulation Parameters [15]

| Simulation Parameter | Value |
|---|---|
| Number of subcarrier (N) | 256 |
| Cyclic prefix length ($N_{cp}$) | 32 |
| Sampling interval ($T_s$) | 250 nsec |
| Photodetector responsivity (r) | $0.28\ ^A/_W$ |
| Process delay ($T_p$) | 50 nsec |
| Pulse shaping filters ($g_T(t)$ and $g_R(t)$) | Square root raised cosine filter with roll-off factor 0.5 |
| Noise power spectral density | $10^{-20}\ ^W/_{H_z}$ |

In Fig 3, we are comparing the HD relay assisted transmission with EPA and OPA, in EPA $K_p = 0.5$ means the signal power get distributed between source and relay terminal equally and in OPA $K_p = 0.8118$ i.e., optimized value so that relay terminal consumes only 19 % of the total power. From Fig 3, we investigate that BPSK-SIM giving better BER performance with OPA rather than EPA for both 2-PSK and square M-QAM in the order of $10^{-4}$. Further we can improve the performance of a relay assisted system by using FD relaying as shown in Fig 4.

In Fig 4, we have considered the transmission with EPA and OPA. By referring to Fig 4, we can say that the FD relaying completely outperforms the HD relaying for all the three modulation techniques with both EPA and OPA as we reach a BER value of $10^{-12}$. The comparison of BER performance of direct, HD, and FD relaying schemes is given in Table 4.

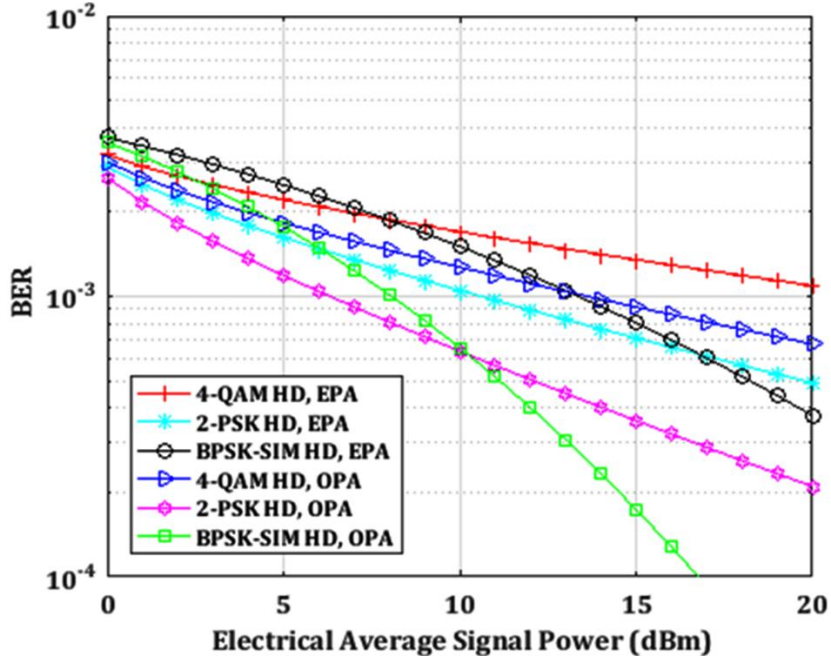

Fig. 3. Comparison of BER performance of half-duplex (HD) relay assisted transmission for EPA and OPA

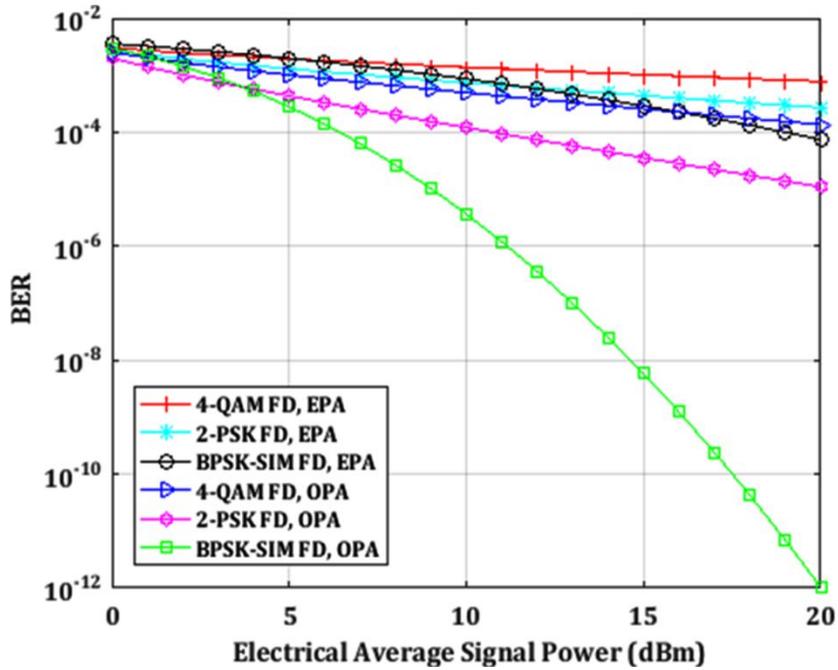

Fig. 4. Comparison of BER performance of full-duplex (FD) relay assisted transmission for EPA and OPA

TABLE 4: Comparison of BER performance of direct, HD and FD systems

| Half Duplex | BER | Full Duplex | BER |
|---|---|---|---|
| Direct 4-QAM | $10^{-1}$ | Direct 4-QAM | $10^{-1}$ |
| Direct 2-PSK | $10^{-2}$ | Direct 2-PSK | $10^{-2}$ |

| Direct BPSK-SIM | $10^{-2}$ | Direct BPSK-SIM | $10^{-2}$ |
|---|---|---|---|
| 4-QAM HD EPA | $2 \times 10^{-3}$ | 4-QAM FD EPA | $10^{-3}$ |
| 2-PSK HD EPA | $8 \times 10^{-4}$ | 2-PSK FD EPA | $10^{-3}$ |
| BPSK-SIM HD EPA | $9 \times 10^{-4}$ | BPSK-SIM FD EPA | $10^{-4}$ |
| 4-QAM HD OPA | $10^{-3}$ | 4-QAM FD OPA | $10^{-3}$ |
| 2-PSK HD OPA | $4 \times 10^{-4}$ | 2-PSK FD OPA | $6 \times 10^{-5}$ |
| BPSK-SIM HD OPA | $2 \times 10^{-4}$ | BPSK-SIM FD OPA | $10^{-8}$ |

From Fig 5 and Fig 6, we illustrate FD relaying and HD relaying through EPA and OPA. We obtain better BER performance over BPSK-SIM with FD relaying scheme irrespective of EPA or OPA in comparison to the other two modulation techniques i.e., 2-PSK and square M-QAM. It is also illustrated that through OPA we achieve better BER than EPA which utilizes FD and HD relaying.

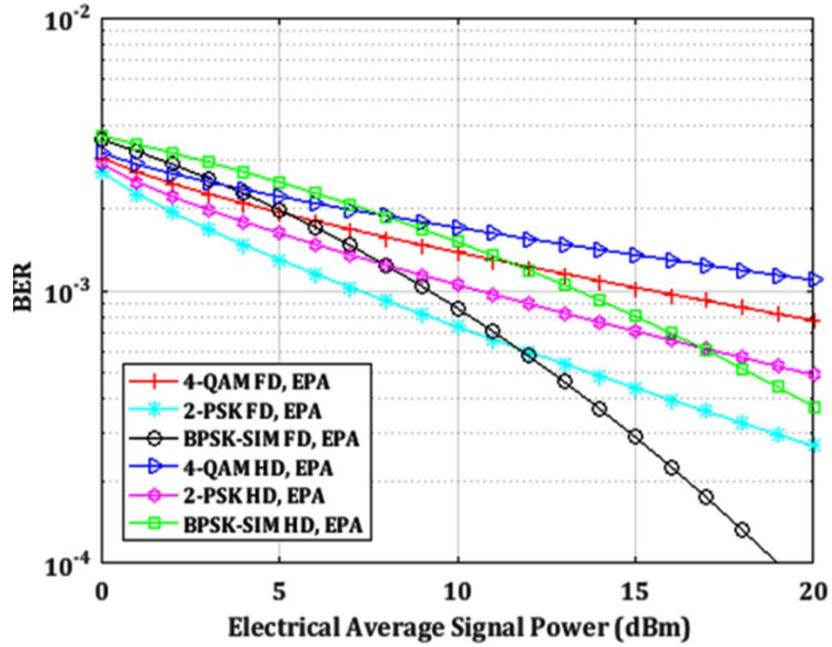

Fig 5. BER performance analysis with EPA

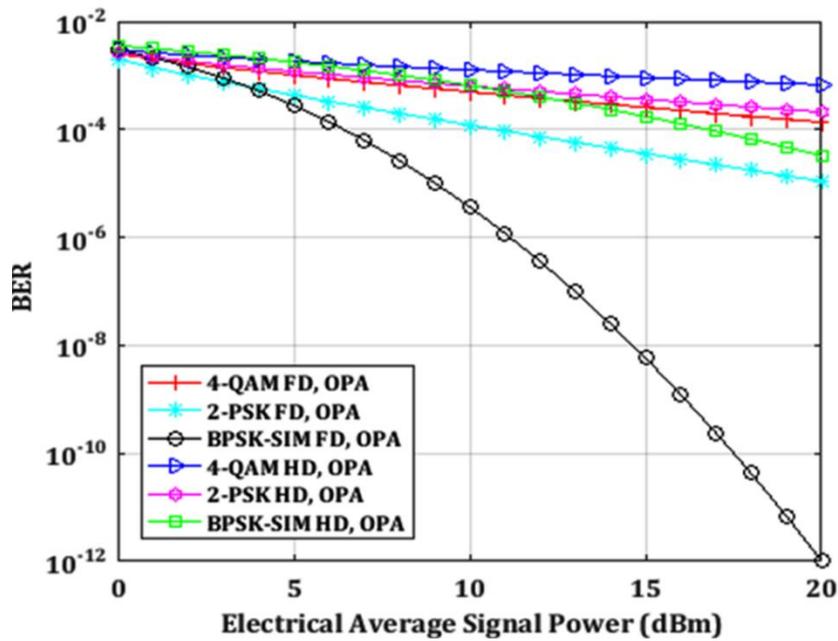

Fig 6. BER performance analysis with OPA

The BER performance for BPSK-SIM through OPA is compared with EPA for both FD and HD relaying is illustrated in Fig 7. BPSK-SIM HD with OPA we obtain a BER range of $10^{-4}$ and in case of BPSK-SIM FD with OPA we achieve a BER value of $10^{-12}$. From figure, it can be noticed that FD relaying benefits atleast 10 dB from EPA to OPA at $10^{-2}$ BER, and it completely outperform the HD relaying with respect to FD relaying with EPA and OPA.

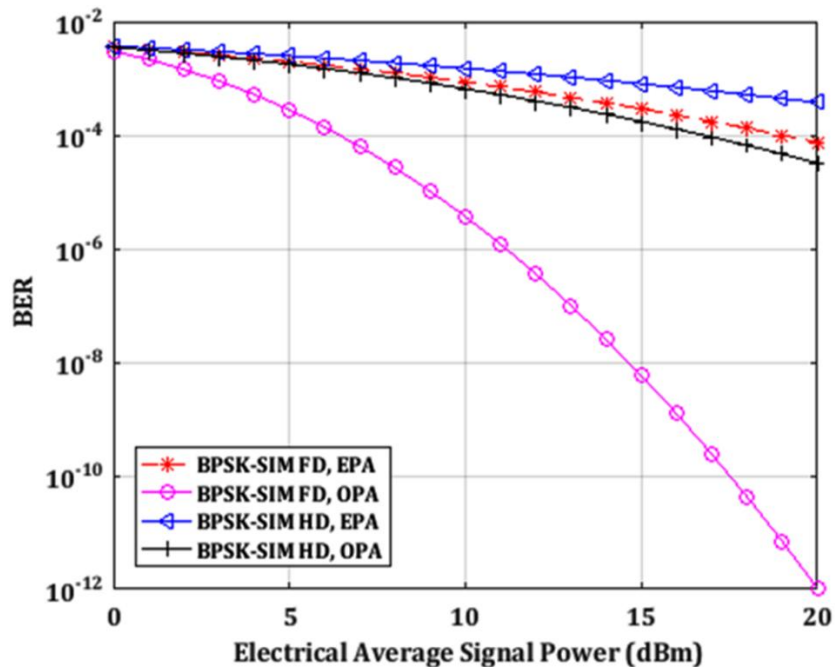

Fig 7. Comparison of BER performance for BPSK-SIM with EPA and OPA

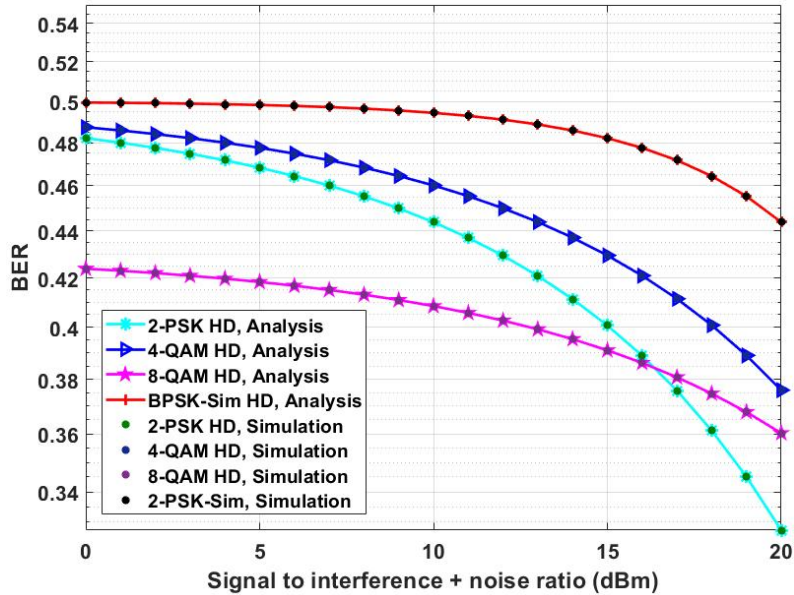

Fig 8. BER Vs SINR performance for different modulation schemes

Figure 8 portrays the BER performance over different modulation schemes for varying signal to interference plus noise ratio (SINR). The results show that the 8-QAM scheme gives good performance over the BPSK and 4-QAM schemes. As the SINR increases the 2-PSK scheme gives a better performance than the other aforementioned modulation schemes.

## 6. Conclusions

In this paper, we have explored the performance of the FD relaying VLC system and compared this with HD relaying VLC system. First, we have considered the office space with ceiling light as source terminal and desk light as a relay terminal. Our results reveal that the FD relaying completely outperform the HD relaying through EPA, also we have investigated the BER performance that can be further enhanced through OPA. We also present that BPSK-SIM gives better performance than traditional 2-PSK and square M-QAM schemes. With the OPA we have achieved a BER value of $10^{-12}$ for BPSK-SIM which is far better than other two modulation schemes.

**Declarations**

**Ethical Approval :** Not applicable

**Competing interests :** Not applicable

**Authors' contributions :** Main manuscript text - Datta Choudhari, Bhargava Kumar L

Figures preparation - Ramavath Prasad Naik, Goutham Simha GD

Derivation and Coding - Datta Choudhari, Bhargava Kumar L

Reviewed Manuscript - Prabu Krishnan, Jagadeesh Vellakudiyan

**Funding :**  Not applicable

**Availability of data and materials :**  Not applicable

**Reference**

[1] Uysal, M., Capsoni, C., Ghassemlooy, Z., Boucouvalas, A., &Udvary, E. (Eds.). *Optical wireless communications: an emerging technology*. Springer, 2016.

[2] Tiwari, Samrat Vikramaditya, Atul Sewaiwar, and Yeon-Ho Chung. "An efficient repeater assisted visible light communication." In *European Wireless 2015; 21th European Wireless Conference; Proceedings of*, pp. 1-5. VDE, 2015.

[3] Yang, Hongming, and Ashish Pandharipande. "Full-duplex relay VLC in LED lighting linear system topology." In *Industrial Electronics Society, IECON 2013-39th Annual Conference of the IEEE*, pp. 6075-6080. IEEE, 2013.

[4] Yang, Hongming, and Ashish Pandharipande. "Full-duplex relay VLC in LED lighting triangular system topology." In *Communications, Control and Signal Processing (ISCCSP), 2014 6th International Symposium on*, pp. 85-88. IEEE, 2014.

[5] Riihonen, Taneli, Stefan Werner, and RistoWichman. "Hybrid full-duplex/half-duplex relaying with transmit power adaptation." *IEEE Transactions on Wireless Communications*10, no. 9 (2011): 3074-3085.

[6] Komine, Toshihiko, and Masao Nakagawa. "Fundamental analysis for visible-light communication system using LED lights." *IEEE transactions on Consumer Electronics* 50.1 (2004): 100-107.

[7] Miramirkhani, Farshad, and Murat Uysal. "Channel modeling and characterization for visible light communications." *IEEE Photonics Journal* 7, no. 6 (2015): 1-16.

[8] Kizilirmak, RefikCaglar, Omer Narmanlioglu, and Murat Uysal. "Relay-assisted OFDM-based visible light communications." *IEEE Transactions on Communications* 63.10 (2015): 3765-3778.

[9] Kazemi, H. and Haas, H., 2016, May. Downlink cooperation with fractional frequency reuse in DCO-OFDMA optical attocell networks. In *Communications (ICC), 2016 IEEE International Conference on* (pp. 1-6). IEEE.

[10] Uysal, Murat, TuncerBaykas, Farshad Miramirkhani, Nikola Serafimovski, and Volker Jungnickel. *TG7r1 channel model document for high rate PD communications*. 2015.

[11] Uysal, Murat, Farshad Miramirkhani, Omer Narmanlioglu, TuncerBaykas, and ErdalPanayirci. "IEEE 802.15. 7r1 reference channel models for visible light communications." *IEEE Communications Magazine* 55, no. 1 (2017): 212-217.

[12] Uysal, M., Miramirkhani, F., Baykas, T., Serafimovski, N. and Jungnickel, J., 2015. *LiFi channel models: office, home, manufacturing cell*. IEEE.

[13] Grobe, Liane, and Klaus-Dieter Langer. "Block-based PAM with frequency domain equalization in visible light communications." In *Globecom Workshops (GC Wkshps), 2013 IEEE*, pp. 1070-1075. IEEE, 2013.

[14] Gonzalez, Gustavo J., Fernando H. Gregorio, Juan Cousseau, TaneliRiihonen, and RistoWichman. "Performance analysis of full-duplex AF relaying with transceiver hardware impairments." In *European Wireless 2016; 22th European Wireless Conference; Proceedings of*, pp. 1-5. VDE, 2016.

[15] Narmanlioglu, Omer, RefikCaglarKizilirmak, Farshad Miramirkhani, and Murat Uysal. "Cooperative visible light communications with full-duplex relaying." *IEEE Photonics Journal* 9, no. 3 (2017): 1-11.

[16]    Narmanlioglu, Omer, Gerard Djengomemgoto, and Murat Uysal. "Performance analysis and optimization of unipolar OFDM based relay-assisted visible light communications☆." *Optik-International Journal for Light and Electron Optics* 151 (2017): 77-87.

[17]    Cho, Kyongkuk, and Dongweon Yoon. "On the general BER expression of one-and two-dimensional amplitude modulations." *IEEE Transactions on Communications* 50, no. 7 (2002): 1074-1080.

[18] Chatterjee, Sourish, Deblina Sabui, Gufran S. Khan, and Biswanath Roy. "Signal to interference plus noise ratio improvement of a multi-cell indoor visible light communication system through optimal parameter selection complying lighting constraints." Transactions on Emerging Telecommunications Technologies 32, no. 10 (2021): e4291.

[19] Z. Wang, Y. Liu, Y. Lin, and S. Huang, ''Full-duplex MAC protocol based on adaptive contention window for visible light communication,'' IEEE/OSA J. Opt. Commun. Netw., vol. 7, no. 3, pp. 164–171, Mar. 2015.